\documentclass[11pt,a4paper]{article}
\usepackage{amssymb} \usepackage{amsmath} \usepackage{graphicx}
\usepackage{epsfig,latexsym}
\begin{document}

\def\bef{\begin{figure}}
\def\eef{\end{figure}}
\newcommand{\ans}{ansatz }
\newcommand{\be}[1]{\begin{equation}\label{#1}}
\newcommand{\beq}{\begin{equation}}
\newcommand{\ee}{\end{equation}}
\newcommand{\beqn}[1]{\begin{eqnarray}\label{#1}}
\newcommand{\eeqn}{\end{eqnarray}}
\newcommand{\bd}{\begin{displaymath}}
\newcommand{\ed}{\end{displaymath}}
\newcommand{\mat}[4]{\left(\begin{array}{cc}{#1}&{#2}\\{#3}&{#4}
\end{array}\right)}
\newcommand{\matr}[9]{\left(\begin{array}{ccc}{#1}&{#2}&{#3}\\
{#4}&{#5}&{#6}\\{#7}&{#8}&{#9}\end{array}\right)}
\newcommand{\matrr}[6]{\left(\begin{array}{cc}{#1}&{#2}\\
{#3}&{#4}\\{#5}&{#6}\end{array}\right)}
\newcommand{\cvb}[3]{#1^{#2}_{#3}}
\def\lsim{\raise0.3ex\hbox{$\;<$\kern-0.75em\raise-1.1ex
e\hbox{$\sim\;$}}}
\def\gsim{\raise0.3ex\hbox{$\;>$\kern-0.75em\raise-1.1ex
\hbox{$\sim\;$}}}
\def\abs#1{\left| #1\right|}
\def\simlt{\mathrel{\lower2.5pt\vbox{\lineskip=0pt\baselineskip=0pt
           \hbox{$<$}\hbox{$\sim$}}}}
\def\simgt{\mathrel{\lower2.5pt\vbox{\lineskip=0pt\baselineskip=0pt
           \hbox{$>$}\hbox{$\sim$}}}}
\def\unity{{\hbox{1\kern-.8mm l}}}
\newcommand{\eps}{\varepsilon}
\def\ep{\epsilon}
\def\ga{\gamma}
\def\Ga{\Gamma}
\def\om{\omega}
\def\omp{{\omega^\prime}}
\def\Om{\Omega}
\def\la{\lambda}
\def\La{\Lambda}
\def\al{\alpha}
\newcommand{\ov}{\overline}
\renewcommand{\to}{\rightarrow}
\renewcommand{\vec}[1]{\mathbf{#1}}
\newcommand{\vect}[1]{\mbox{\boldmath$#1$}}
\def\tm{{\widetilde{m}}}
\def\mcirc{{\stackrel{o}{m}}}
\newcommand{\Dm}{\Delta m}
\newcommand{\dm}{\varepsilon}
\newcommand{\tanb}{\tan\beta}
\newcommand{\nbar}{\tilde{n}}
\newcommand\PM[1]{\begin{pmatrix}#1\end{pmatrix}}
\newcommand{\up}{\uparrow}
\newcommand{\down}{\downarrow}
\def\omE{\omega_{\rm Ter}}
%

\newcommand{\Dsusy}{{susy \hspace{-9.4pt} \slash}\;}
\newcommand{\DCP}{{CP \hspace{-7.4pt} \slash}\;}
\newcommand{\mc}{\mathcal}
\newcommand{\gr}{\mathbf}
\renewcommand{\to}{\rightarrow}
\newcommand{\gtc}{\mathfrak}
\newcommand{\wh}{\widehat}
\newcommand{\br}{\langle}
\newcommand{\kt}{\rangle}


\def\lsim{\mathrel{\mathop  {\hbox{\lower0.5ex\hbox{$\sim$}
\kern-0.8em\lower-0.7ex\hbox{$<$}}}}}
\def\gsim{\mathrel{\mathop  {\hbox{\lower0.5ex\hbox{$\sim$}
\kern-0.8em\lower-0.7ex\hbox{$>$}}}}}

\def\nn{\\  \nonumber}
\def\de{\partial}
\def\brf{{\mathbf f}}
\def\bbf{\bar{\bf f}}
\def\bF{{\bf F}}
\def\bbF{\bar{\bf F}}
\def\bA{{\mathbf A}}
\def\bB{{\mathbf B}}
\def\bG{{\mathbf G}}
\def\bI{{\mathbf I}}
\def\bM{{\mathbf M}}
\def\bY{{\mathbf Y}}
\def\bX{{\mathbf X}}
\def\bS{{\mathbf S}}
\def\bb{{\mathbf b}}
\def\bh{{\mathbf h}}
\def\bg{{\mathbf g}}
\def\bla{{\mathbf \la}}
\def\bmu{\mathbf m }
\def\by{{\mathbf y}}
\def\bmu{\mbox{\boldmath $\mu$} }
\def\bsig{\mbox{\boldmath $\sigma$} }
\def\bunity{{\mathbf 1}}
\def\cA{{\cal A}}
\def\cB{{\cal B}}
\def\cC{{\cal C}}
\def\cD{{\cal D}}
\def\cF{{\cal F}}
\def\cG{{\cal G}}
\def\cH{{\cal H}}
\def\cI{{\cal I}}
\def\cL{{\cal L}}
\def\cN{{\cal N}}
\def\cM{{\cal M}}
\def\cO{{\cal O}}
\def\cR{{\cal R}}
\def\cS{{\cal S}}
\def\cT{{\cal T}}
\def\eV{{\rm eV}}
%




\large
 \begin{center}
 {\Large \bf Neutron-Antineutron transitions from exotic instantons: how fast they might be and further implications}
 \end{center}

 \vspace{0.1cm}

 \vspace{0.1cm}
 \begin{center}
{\large Andrea Addazi}\footnote{E-mail: \,  andrea.addazi@infn.lngs.it} \\
{\it \it Dipartimento di Fisica,
 Universit\`a di L'Aquila, 67010 Coppito, AQ \\
LNGS, Laboratori Nazionali del Gran Sasso, 67010 Assergi AQ, Italy}
\end{center}

\vspace{1cm}
\begin{abstract}
\large
We discuss implications of open string theory in B-violating low energy physics. 
In particular, exotic instantons can dynamically generate effective six quarks' operator, 
leading to a neutron-antineutron transition. Proton is not destabilized
and flavor changing neutral currents are under control.

\end{abstract}

\baselineskip = 20pt

\section{Introduction and conclusions}
Recently, we have suggested that exotic stringy instantons 
can have important implications in low energy particle physics.
In particular, we have suggested several explicit models 
generating a Majorana mass for the neutron \cite{Addazi:2014ila,Addazi:2015ata,Addazi:2015rwa,Addazi:2015hka,Addazi:2015eca,Addazi:2015fua,Addazi:2015oba,Addazi:2015goa,Addazi:2015yna}
. Such a signature can be tested in the next generation of experiments in neutron-antineutron physics
improving the present limit of $\tau\simeq 3\, \rm yrs$ (in vacuum), 
up to $\tau \simeq 300\, \rm yrs\,\,\,$
\cite{Phillips:2014fgb}. So fast a transition in vacuum does not contradict 
limits from nuclei destabilization. In fact,  
in Nuclei, the binding energy of neutron will suppress the transition time up to $10^{32}\, \rm yrs$, 
contrary to decays (like proton decays). 
A neutron-antineutron transition can be generated by effective six quarks operators
like $(u^{c}d^{c}d^{c})^{2}/\mathcal{M}^{5}$, 
{\it i.e} by an effective Majorana mass for neutron of $\delta m=\tau^{-1} \simeq \Lambda_{QCD}^{6}/\mathcal{M}^{5}$.
So that, present best limits on $\tau$ constrain $\mathcal{M}$ to be higher than $300\, \rm TeV$
and $\delta m<10^{-23}\, \rm eV$. 
The next generation of experiments will test $\mathcal{M}\simeq  1000\, \rm TeV$. 

Exotic instantons have a very intuitive geometric interpretation: 
they are nothing but Euclidean D-branes, wrapping the Calabi-Yau compactification 
with n-cycles. Exotic instantons can intersect ordinary $(D+4)$-branes, 
   generating new effective interactions among ordinary fields in the low energy limit. 
The most radical suggestion was proposed and studied in Ref.\cite{Addazi:2015goa}: 
we have demonstrated that a Majorana mass term for the neutron can be directly 
generated with one and only one exotic instanton, without the needing of extra matter fields. 
This is a new different and calculable example of an UV completion of an effective non-renormalizzable 
operator: the effective operator is completed by the formation of a non-perturbative classical configuration 
rather than by new heavy fields. 
In this case, future high energy proton-proton colliders can be directly tested the 
resonant production of exotic instantons in collisions:
$qq\rightarrow \bar{q}\bar{q}\bar{q}\bar{q}$ or $qq\rightarrow \bar{\tilde{q}}\bar{\tilde{q}}\bar{\tilde{q}}\bar{\tilde{q}}$
are 'golden channels'. 
In fact, contrary to other models, the associated scattering amplitude is expected to have a 
peculiar behavior that cannot be reproduced by gauge theories. 
In the limit of $\sqrt{s}<<\Lambda$, where $\Lambda$ is the characteristic effective scale associated to the exotic instanton
and supersymmetric reductions, 
amplitudes are expected to be point-like, corresponding to polynomially increasing cross sections with CM energy.
In the limit of momenta much higher than quark masses, 
 $\sigma_{qq\rightarrow \bar{q}\bar{q}\bar{q}\bar{q}} \sim s^{4}/\Lambda^{10}$ 
 for $\sqrt{s}<\Lambda$. 
For $\sqrt{s}\simeq \Lambda$ the cross section is cutoff 
as $\sigma \simeq \frac{1}{\Lambda^{2}}$ \footnote{For $qq\rightarrow \bar{\tilde{q}}\bar{\tilde{q}}\bar{\tilde{q}}\bar{\tilde{q}}$
the cross section is $\sigma \sim \Lambda^{-10}M_{SUSY}^{4}s^{2}$ in the limit of $\sqrt{s}>>m_{\tilde{q}}$. 
The massless limit is physically sensed for $1\, \rm TeV$ supersymmetry scale. 
}
\footnote{Otherwise, causality and unitarity would be violated as usually happen in non-local quantum field theories
\cite{Addazi:2015dxa,Addazi:2015ppa}.}. 
At that scale, also Regge stringy resonances are expected to be produced. 
We are in the fully non-perturbative regime, but for $\sqrt{s}>\Lambda$ the amplitude is expected to be 
exponentially decreasing with Mandelstam variables (not polynomially as in quantum field theories), as usually happen in string theory, 
{\it i.e} unitarity is restored. 
In the limit of $\sqrt{s}>>\Lambda$, one can conjecture a {\it duality} of such an E-brane instanton with 
a worldsheet instanton in the heterotic string theory in the limit of $\sqrt{s}<<\Lambda$.
To demonstrate such a duality in our realistic model is absolutely non-trivial, but 
it seems to be supported by some simpler examples in literature 
\cite{Kiritsis:2000zi,Bianchi:2007rb}. 
Such a duality would allow to calculate perturbatively the scattering amplitude in the limit 
of $\sqrt{s}>>\Lambda$ from the heterotic stringy amplitudes. 
Of course such calculations remain complicate ones, 
but by virtue of this conjecture one could expect a kinematical dependence of the amplitude 
with respect to the scattering amplitude as a sort of generalization of Veneziano's amplitude.
Let us remind that Veneziano's amplitude has a form 
$\mathcal{A}_{V}\sim \Gamma(1-\alpha_{s}s)\Gamma(1-\alpha_{s}u)/\Gamma(2-\alpha_{s}s-\alpha_{s}u)$.
Clearly, in the case of six open stringy amplitudes, a more complicated combination of Gamma functions with more kinematic invariants $s_{ij}=(p_{i}+p_{j})^{2}$. 
But essentially, one can expect again a polynomially increasing amplitude in the limit $\sqrt{s}<<\Lambda$ in the heterotic side (it means $\sqrt{s}>>\Lambda$ in the 
open string side). In this case, an exotic instanton will be found if the stringy scale is $M_{S}\simeq 10\div 10^{5}\, \rm TeV$, 
{\it i.e} in low scale string theory \footnote{However, if $\Lambda\simeq M_{S} \simeq M_{Pl}$, the effect of gravity starts to be relevant 
and the formation of an exotic instanton could be suppressed by the formation of  micro black holes. In this case, 
a black hole could unitarize the effective six quark operator rather than an exotic instanton. 
}. 

On the other hand, we have also suggested other models in which six quarks effective operator are completed 
from exotic instantons and extra colored fields. Also in these case several channels can be tested in the next generation of experiments. 
In particular, a vector-like pair od colored triplets can be found by LHC or future colliders in 
"clean" channels like $pp\rightarrow 4j$ or $pp\rightarrow jjE_{T}$ \cite{Addazi:2015ata,Addazi:2015rwa}. These models have also the advantage to connect 
neutron-antineutron transitions to Post-Sphaleron baryogenesis, {\it i.e} to a first order phase transition \cite{Addazi:2015ata}. 
In these indirect mechanisms, we did not need to consider a low scale string theory scenario, even if compatible with it. 

As a consequence, we can anticipate the answer to the issue raised in our title:
limits on neutron-antineutron transition from exotic instantons are directly bounded by 
neutron-antineutron oscillations' experiments in vacuum, 
without other stronger indirect constraints (like proton decays, flavor changing neutral currents and so on). 
This strongly motivates future experiments in neutron-antineutron transition as an indirect
test-bed for several different model with exotic stringy instantons. 
An then, as partially mentioned above and fully discussed in references cited above, several different
channels can test and constrain our models. 

In the next section, we will summarize the main technical aspects of our proposal. 
In particular, we will see how the standard model of particles can be UV completed 
in open string theories. We will see how in these models a neutron Majorana mass can be 
easily generated by calculable and controllable mixed disk amplitudes. 
Finally, we will discuss phenomenological implications in Section 3. 

\section{UV completion of the Standard Model and exotic instantons}

The SM can be UV completed in IIA open string theory, considering a system of intersecting D6-branes' stacks. 
$SU(3)_{c}\subset U(3)_{c}$ can be embedded in a system of three parallel D6-branes, 
as well as $SU(2)_{L} \subset Sp(2)_{L}$ in a stack of two D6-branes and so on. 
In Fig.1-(b), a system of intersecting D6-branes and open strings 
UV completing the (MS)SM is shown. Similar diagrams are called quivers. 
A quiver can encode all informations about the dynamics in the low energy limit:
all fundamental fields, interactions and quantum consistency.
To be more precise about the last point, a quiver encodes all informations 
about anomaly cancellations nearby the Calabi-Yau singularity
in which D-branes' stacks are located. 
In the low energy limit, 
D-branes are practically non dynamical entities, while 
SM superfields are obtained as lowest excitations of the open (un)oriented strings 
attached to the D6-branes' stacks \footnote{The Dp-brane tension is $T_{Dp}\sim 1/(g_{s}\alpha_{s}^{(p+1)/2})$, 
with dimension in mass $[T_{Dp}]=M^{p+1}$ ($\alpha_{s}=l_{s}^{2}$). 
The decoupling of gravity is in the limit of $g_{s}\rightarrow 0$, corresponding to $T_{Dp}\rightarrow \infty$ (D-branes
becomes rigid "walls"). In fact, the coupling of D-branes with gravity is $\kappa^{2} T'\sim g_{s}^{4}/g_{s}$, 
so that the gravitational back-reaction can be neglected for $g_{s}\rightarrow 0$. 
The Dp-brane tension depends with $g_{s}$, contrary to string tension
 $T'=(\alpha_{s}/2\pi)^{-1}$. Massive Regge and KK modes are decoupling in the limit of 
 $l_{s}=\sqrt{\alpha'}\rightarrow 0$ and $R\rightarrow 0$ (KK R compactification radius)}. 
The convention used in a quiver are the following: the black nodes correspond to D6-branes' stacks 
reproducing gauge groups; (un)oriented arrows are associated to (un)oriented open strings attached to 
D6-branes' stacks, {\it i.e} they correspond to superfields in the bifundamental rapresentations 
of gauge groups; while the number of arrows correspond to the number of intersections among the 
D6-branes' stacks, {\it i.e} to the number of generations of each superfields. 
To give a concrete example, the quark doublet superfields $Q$ correspond to three arrows between 
the two nodes ${\bf 3}$ and ${\bf 2}$ in Fig.1-(b); {\it i.e} they are obtained as the lowest excitations of open strings attached 
to a stack of three D6-branes and a stack of two D6-brane on the $\Omega$-plane. In particular, these stacks are intersecting 
three times in order to recover the correct number of (super)quarks' generations. 

The low energy theory is $\mathcal{N}=1$ supersymmetric gauge one 
with a gauge symmetry 
$$U(3)_{c}\times Sp(2)_{L}\times U(1) \times U(1)' \rightarrow SU(3)_{c}\times Sp(2) \times U(1)_{Y}$$
where the two extra anomalous $U(1)s$ are broken through a Stueckelberg mechanism, 
while $U(1)_{Y}$ is a non-anomalous massless combination of $U(1)_{3},U(1),U(1)'$ 
$$U(1)_{Y}=\frac{1}{3}U(1)_{3}-U(1)+U(1)'$$
In gauge theories anomalous gauge $U(1)$s cannot be consistently considered. 
However, in string theory, they can be cured through a generalized Green-Schwarz 
mechanism if the number of intersections among $D6$-branes' stacks satisfy certain 
conditions for stringy tadpoles' cancellations, as happen in quiver Fig.1-(b) \cite{Addazi:2015rwa}. 
This mechanism necessary involves extra Generalized Chern-Simons terms,
mixing at tree level the anomalous $Z',Z''$, associated with anomalous $U(1)$s, 
with neutral SM bosons $Z,\gamma$. A generalized Chern-Simons term is 
$\epsilon_{\mu\nu\rho\sigma}F_{(1)}^{\mu\nu}A_{(2)}^{\rho}A_{(3)}^{\sigma}$, 
with $1,2,3$ label three neutral bosons among $Z',Z'',Z,\gamma$. 
The matter superfields obtained in the low energy limits are $Q,U^{c},D^{c},L,E^{c},N^{c},H_{u},H_{d}$.
The superpotentials allowed at perturbative level correspond to closed circuit triangles 
with the same orientation of arrows. In quiver Fig.1-(b), Yukawa's superpotentials are reproduced at perturbative level: 
$$\mathcal{W}_{p}=y_{E}H_{d}LE^{c}+y_{N}H_{u}^{\alpha}L_{\alpha}N^{c}+y_{U}H_{u}QU^{c}+y_{D}H_{d}QD^{c}$$
while the mu-term $\mu H_{u}H_{d}$ can be generated by R-R or NS-NS fluxes. 

The E2-instanton in Fig.1-(b) has a Chan-Paton group $O(1)$.
It generates the following effective interactions
in the low energy limit:
$$\mathcal{L}_{eff}\sim C_{f}^{(1)}U^{i}_{f}\tau_{i}\alpha+C_{f}^{(2)}D^{i}_{f}\tau_{i}\beta$$
where $\tau_{i}$ are modulini coming from $U(3)-E2$ (two) intersections, 
$\alpha$ modulini from $U(1)-E2$ (two) intersections
and $\beta$ moduli from $U(1)'-E2$ (four) interesections. 
As a consequence an effective superpotential 
$$\mathcal{W}_{E2}=\int d^{6}\tau d^{4}\beta d^{2}\alpha e^{\mathcal{L}_{eff}}$$
$$=\frac{e^{-S_{E2}}}{M_{S}^{3}}C_{f1}^{(1)}C_{f2}^{(2)}C_{f3}^{(2)}C^{(1)}_{f4}C^{(2)}_{f5}C^{(2)}_{f6}
\epsilon_{ijk}\epsilon_{i'j'k'}U_{R}^{i,f1}D_{R}^{j,f2}D_{R}^{k,f3}U_{R}^{i',f4}D_{R}^{j',f5}D_{R}^{k',f6}$$
after integration over the modulini 
$e^{+S_{E2}}$ is a function of the geometric moduli associated to the 3-cycles wrapped by the E2-instanton. 
In a local model, this can be considered as a free parameter.

\begin{figure}[t]
\centerline{\includegraphics [height=9cm,width=1\columnwidth] {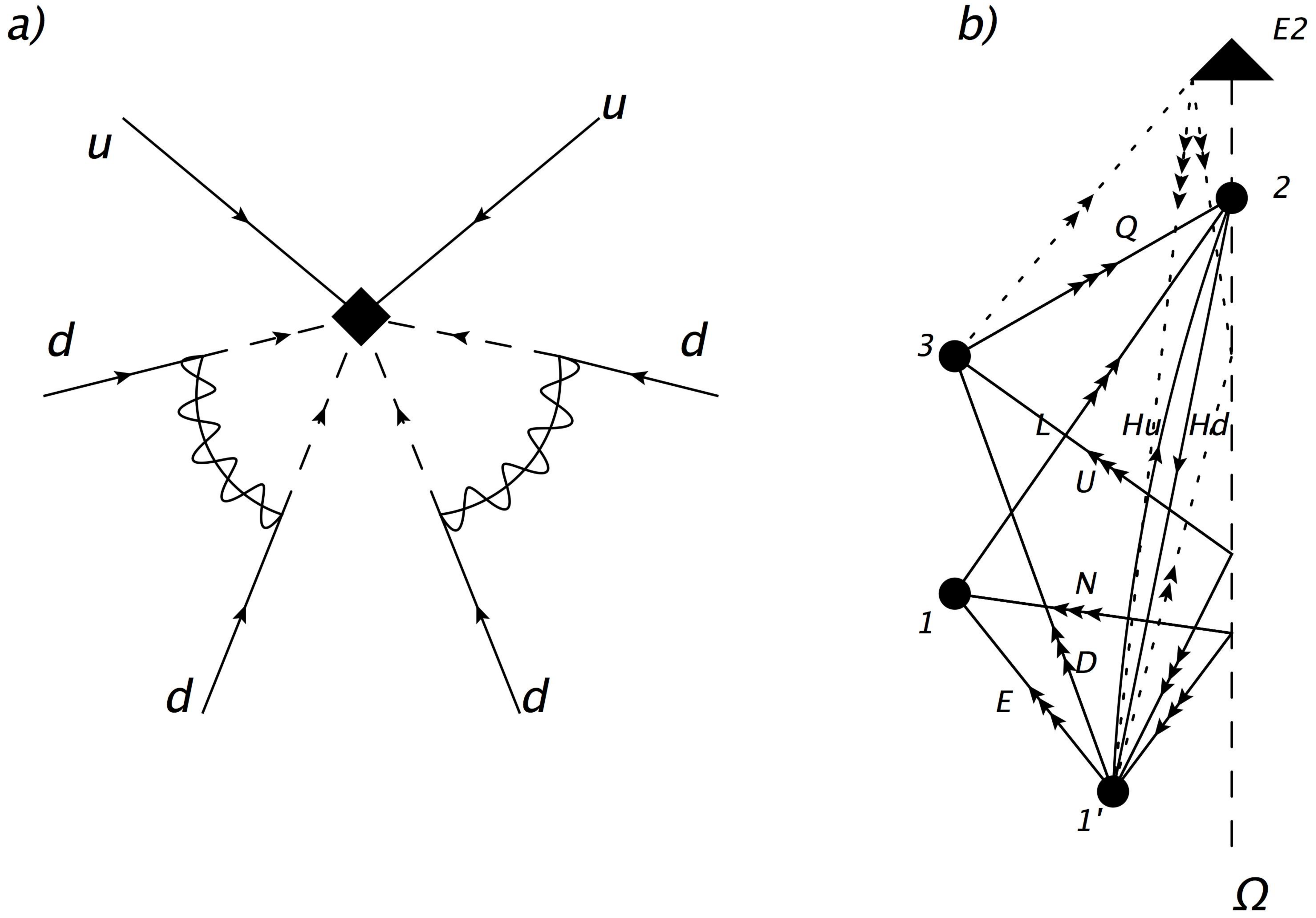}}
\vspace*{-1ex}
\caption{In (a) a neutron-antineutron diagram induced by exotic instantons. In particular a superpotential is non-perturbatively generated so that 
a contact interaction among two squarks and one quark is dynamically introduced. In (b) we show a possible unoriented quiver theory embedding the SM. 
The relevant exotic instanton is represented as a triangle. 
}
\label{plot}   
\end{figure}

\section{Phenomenology in Neutron-Antineutron transitions and further implications}

The superpotential generated in section 2 contains 
a neutron-antineutron operator. 
After supersymmetric reductions, diagramatically shown in Fig.1-(a), 
we obtain the operator in the lagrangian density
$$\mathcal{O}_{n\bar{n}}=\frac{\mathcal{Y}}{\mathcal{M}^{5}}(u^{c}d^{c}d^{c})^{2}$$
where 
$$\mathcal{Y}=C_{1}^{(1)}C_{1}^{(2)}C_{1}^{(2)}C^{(1)}_{1}C^{(2)}_{1}C^{(2)}_{1}$$
and 
$$\mathcal{M}^{5}=m_{\tilde{g}}^{2}M_{S}^{3}e^{+S_{E2}}$$
where $m_{\tilde{g}}$ is the gaugino mass (gluino, zino or photino). 
In order to generate a Majorana mass term for neutron of the order 
of $\delta m\simeq 10^{-25}\, \rm eV$, testable in the next generation of experiments, 
$\mathcal{M}\simeq 1\, \rm PeV$. 
This scale can be recovered in several different regions of parameters 
$(m_{\tilde{g}},M_{S},e^{+S_{E2}},\mathcal{Y})$. 
For example, for $m_{\tilde{g}}\simeq M_{S}\simeq 1\, \rm PeV$
and $\mathcal{Y}\simeq e^{+S_{E2}}\simeq 1$
can recover the desired PeV-scale. 
Geometrically, an $e^{+S_{E2}}$ corresponds to a very small size 3-cycles wrapped 
by the $E2$-instanton on the Calabi-Yau. 
On the other hand, we can also envisage a scenario with TeV-scale supersymmetry breaking
in which $m_{\tilde{g}}\simeq 1\div 10\, \rm TeV$, $M_{S}\simeq 10^{5}\, \rm TeV$
and $\mathcal{Y}\simeq e^{+S_{E2}}\simeq 1$. 
Finally, another interesting region of parameters corresponds to 
$m_{\tilde{g}}\simeq M_{S}\simeq 10\, \rm TeV$, $\mathcal{Y}\simeq 1$ 
and $e^{+S_{E2}}\simeq 10^{10}$ corresponding to very large 3-cycles wrapped by the exotic instantons
\footnote{Alternatively one can redefine $e^{+S_{E2}}$ as $e^{+\tilde{S}_{E2}}=e^{+S_{E2}/5}$, 
so that $e^{+\tilde{S}_{E2}}\simeq 100$ for $e^{S_{E2}}\simeq 10^{10}$.}
This last case is the most exciting one for LHC, that could directly detect stringy Regge resonances 
and anomalous $Z'$-bosons, as well as supersymmetric partners. 

Let us note that in our model the proton is not destabilized, as discussed in more details in 
\cite{Addazi:2015fua,Addazi:2015goa}. In fact, our mechanism generates 
a $\Delta B=2$ violating operator, while $\Delta L=1$ operators are not generated at all, 
{\it i.e} all dangerous proton decay channels are avoided. 
In other words, R-parity is dynamically broken so that one has a new residual selection rule 
$\Delta B=2$. As discussed in \cite{Addazi:2015fua,Addazi:2015goa}, not even other rare processes 
can competitively constrain the six quark operator generated in our model.

Finally, let us also note that other six quarks operators with all possible combinations of flavors 
are generated in our model: $uds\rightarrow \bar{u}\bar{d}\bar{s}$,
$cbs\rightarrow \bar{c}\bar{b}\bar{s}$, and so on.
This can have important implications for future colliders of $E_{CM}\sim 100\, \rm TeV$:
they could directly detect $\Delta B=2$ six quarks' collisions 
Infact, the scale $e^{+S_{E2}}(M_{S}^{3}m_{\tilde{g}}^{2})$ can be as small as $100\, \rm TeV$ if $\mathcal{Y}\simeq 10^{-5}$
while for example 
$C_{2}^{(1)}C_{2}^{(2)}C_{2}^{(2)}C^{(1)}_{2}C^{(2)}_{2}C^{(2)}_{2}\simeq 1$.
In this case, the effective new physics scale $\mathcal{M}$ of $n-\bar{n}$ 
remains $1\, \rm PeV$ or so, but 
the ones of different flavors can be smaller than $1\, \rm PeV$ by a factor $10^{-1}$.
The possibility to directly detect exotic instantons in high energy colliders beyond LHC 
would be a spectacular signature of string theory, for the motivations mentioned above 
in Section 1.

\vspace{1cm} 

{\large \bf Acknowledgments} 
\vspace{4mm}

I am particularly grateful to Massimo Bianchi for very interesting discussions and remarks on these subjects.


\end{document}